# Evidence for Extraction of Photoexcited Hot Carriers from Graphene

*Chang-Hua Liu, Nanditha M. Dissanayake, Seunghyun Lee, Kyunghoon Lee and Zhaohui Zhong\**

Department of Electrical Engineering and Computer Science, University of Michigan, Ann Arbor, Michigan 48109

\*Corresponding author. Electronic mail: zzhong@umich.edu

ABSTRACT

We report evidence of nonequilibrium hot carriers extraction from graphene by gate-dependent photocurrent study. Scanning photocurrent excited by femtosecond pulse laser shows unusual gate dependence compared with continuous wave (CW) laser excitation. Power dependence studies further confirm that the photocarriers extracted at the metal/graphene contact are nonequilibrium hot carriers. Hot carrier extraction is found to be most efficient near the Dirac point where carrier lifetime reaches maximum. These observations not only provide evidence of hot carrier extraction from graphene, but also open the door for graphene based hot carrier optoelectronics.

Graphene, composed of single layer carbon atoms, could be the potential material for hot carrier optoelectronic applications. The strong optical absorption of graphene (2.3% for a single atomic layer)[1-2] and the reduced phonon modes in low dimension indicate the possibility of creating nonequilibrium hot carriers due to inefficient cooling when the optical phonon temperature quickly rises to near hot carrier temperature under intense excitation.[3] In addition, as hot carriers cool below optical phonon energy (~200meV), inefficient carrier-acoustic phonon relaxation process can further slow down carrier cooling.[4-5] Recently, hot-carrier dynamics in graphene, including hot-carrier diffusion,[6] carrier-carrier interaction,[7-8] carrier-phonon coupling and carrier recombination,[9-11] were investigated by femtosecond pulse laser. On the other hand, photocarrier transport in graphene, excited by CW laser, was found to follow built-in electric field[12-15] or photo-thermoelectric effect (PTE).[16-18] For the former mechanism, photoexcited electrons and holes are accelerated by built-in electric field originated from the work function difference across the junction. In the latter mechanism, however, photo-thermoelectric current is driven by photocarrier diffusion when temperature gradient is established across interface with different thermal power. Both mechanisms show photocurrent polarity reversal by tuning graphene doping concentration. Although it is still challenging to distinguish the dominant photocurrent generation mechanism,[17, 19] recent studies at graphene pn junction suggest inefficient electron – acoustic phonon relaxation will enhance thermoelectric current, with a signature of multiple polarity reversals.[18, 20-21]

To this end, we report photocurrent studies at graphene-metal junction and graphene pn junction by using both femtosecond pulse laser and CW laser excitation. Surprisingly, the gate dependent photocurrent generated at graphene-metal junction does not exhibit polarity reversal under pulse laser excitation. In addition, photocurrent peaks near the graphene Dirac point gate voltage, where photocarrier lifetime reaches maximum based on theories.[22-24] The results provide the evidence of hot carriers generation and extraction from the graphene device. Also, the mechanism of photocurrent generation within pristine graphene pn junction is confirmed to be due to photo-thermoelectric effect.

RESULTS AND DISCUSSION

Figure 1a shows the schematic of a typical graphene pn junction formed by electrostatic gating. A pair of split bottom gates ($V_{g1}$ and $V_{g2}$) can electrostatically dope the graphene in the above sections into p and n-type, respectively, and form the pn junction in between (details in Supporting Information). Resistance *versus* split gate voltages scans exhibit consistent gate responses for both gates and a Dirac point gate voltage of ~4.8V (Fig. 1b). The device is studied by scanning photocurrent spectroscopy[13] (Figure 1a, also see Supporting Information) at ambient conditions. Briefly, we raster scan the excitation laser across the device, and simultaneously measure the photocurrent and reflected light intensity. Both CW ($\lambda$ = 900 nm) and Femtosecond pulsed laser ($\lambda$ = 800 nm) are used, and the focused laser spot sizes are around 1.5 µm. The upper inset in Fig. 1b shows a spatially resolved photocurrent map with CW excitation under zero source-drain and gate bias voltage. Photocurrent peaks at the source and drain metal-graphene contacts, confirmed by overlapping with the reflected light intensity mapping (Fig. 1b lower inset).

We then turn our attention to the gate dependent photocurrent mapping across the length of the device (dotted line in the upper inset of Fig. 1b) with both CW and pulse laser excitation. Figure 1c and 1d show photocurrent *versus* $V_{g2}$ and laser position for CW and femtosecond pulse excitation, respectively. $V_{g2}$ is scanned from -10 to 20V with $V_{g1}$ grounded during the measurement, modulating the graphene device from pp junction to pn junction. Significantly, two distinct differences are observed by comparing the CW *versus* pulse laser excited photocurrent maps. First, photocurrent peaks at the pn junction in between the split-gates with CW excitation (Figure 1c, Position = 2.5 µm), but disappears when excited with pulse laser (Figure 1d, Position = 2.5 µm). Second, photocurrent near the left metal/graphene junction also shows drastic difference for CW and pulse excitation. Under CW excitation, photocurrent switches sign at $V_{g2}$ = 7.5V (Fig. 1c, Position = 0). Surprisingly, with pulse excitation, photocurrent remains positive, and peaks at $V_{g2}$ = 5V (Fig. 1d, Position = 0). The same phenomena are also observed at the right metal/graphene contact when tuning $V_{g1}$ gate voltages with $V_{g2}$ grounded, and reproducible among all devices tested.

We first focused our attention on the photocurrent abnormally at the metal/graphene contact between CW and pulse laser excitation. Fig. 2a and 2b show the gate dependent photocurrent at the contact edge extracted from Fig. 1c and 1d at Position = 0, respectively. In Fig. 2a, the photocurrent excited by CW laser switches sign at $V_{g2}$ = 7.5 V, agreeing with the literatures where work function difference between graphene and metal determines the sign of photocurrent.[12, 14-15] Based on the thickness of the gate dielectric (50 nm $Al_2O_3$, $\varepsilon$~7.5) and the Dirac point voltage, we estimate a metal work function of 4.3 eV, consistent with the typical value for our contact metal Ti. In comparison, surprisingly, photocurrent excited by the pulse laser (Fig. 2b) remains positive throughout the gate voltage sweep and peaks at $V_{g2}$ = 5 V, which coincides with the Dirac point gate voltage. Furthermore, the photocurrent curve exhibits nearly symmetrical decay around $V_{g2}$ = 5 V by increasing either hole density or electron density.

The unusual photocurrent response from pulse laser excitation provides the evidence of nonequilibrium hot carrier extraction from graphene. Photo-generated hot carriers typically release energy to optical phonons within sub-picosecond time scale, followed by slower relaxation through scattering with acoustic phonons and electron-hole recombination.[4-5] Hot carrier relaxation through carrier multiplication has also been reported in graphene recently.[18, 25] Nevertheless, under CW excitation, photocarriers will relax and accumulate near the equilibrium Fermi level (Fig. 2c), and photocurrent arises from the extraction of these near-equilibrium carriers by either electric field or local temperature gradient.[12-15] However, the lack of photocurrent polarity reversal hints a different mechanism. Under pulse illumination, a high flux of photon excitation will create excess amount of hot carriers. The relaxation of these hot carriers through scattering with optical phonon will quickly raise the optical phonon temperature to near hot carrier temperature, which becomes the bottleneck for thermal relaxation of hot carriers.[3, 9] These processes lead to nonequilibrium hot carriers with elevated quasi Fermi level (Fig. 2d). As a result, hot carrier transport does not follow conventional mechanisms, and hot carrier photocurrent is proportional to its lifetime rather than the metal-graphene built-in electric field. Recent theoretical works predict hot-carrier lifetime will decrease with increasing intrinsic carrier density as Fermi energy moves away from the Dirac point.[22-24] Our observation of peak photocurrent at the Dirac point (Fig. 2b) agrees with the prediction and provides evidence for the nonequilibrium hot carrier induced photocurrent in graphene. The exact mechanism for hot carrier extraction is unclear from this work. However, we speculate that the asymmetric electron and hole mobilities ($\mu_e/\mu_h$=0.86 for the device shown in Fig. 1b) can lead to hot electron and hole diffusion with different velocity. The resulting spatial charge distribution builds up the transient electric field, which drives the carriers to the contact. On-going works on devices with different metal contacts and different electron and hole mobilities are underway to elucidate the hot carrier extraction mechanism.

To gain further insight into the hot carrier dynamics, we studied the gate dependent photocurrent at the metal/graphene junction under different pulse laser power. Figure 3a-c show three representative photocurrent maps measured at 580 μW, 930 μW and 3.49 mW pulse laser power, respectively. Significantly, photocurrent switches sign as the pulse laser power drops (Fig. 3a), reminiscent of the case with CW excitation. More detailed power dependent photocurrent curves obtained at the contact edge are shown in figure 3d with the zoom-in view shown in figure 3e. At a low power of 145 μW, photocurrent switches sign at $V_{g2}$ ~7.5 V and positive photocurrent peaks around 0.5 V. These values are similar to the curve shown in Fig. 2a, suggesting built-in field and PTE dominate current generation. By increasing power to 580 µW, photocurrent amplitude increases in both positive and negative regions. However, at 930 µW, photocurrent becomes entirely positive and peaks at 2V, indicating hot carriers extraction also contribute to photocurrent generation. With further increasing of laser power, positive photocurrent peak gradually shifts to 5 V (Fig. 3e, inset), at which point hot carriers dominate transport. The results indicate that hot carrier extraction is closely related to pulse laser illumination power.

We also studied the gate dependent photocurrent generation within the graphene pn junction formed by the split bottom gates. To identify the photocarrier transport mechanism, we compared the experimental data with simulations of field driven carrier transport and PTE originated transport in figure 4. From the split gate responses of the device, we can calculate gate dependent thermopower (Fig. 4a) using the Mott formula[16]:

$$S = -\frac{\pi^2 K_B^2 T}{3e} \frac{1}{G} \frac{dG}{dE}\bigg|_{E=E_f}$$

where $S$ is thermopower and $G$ is conductance. PTE originated photocurrent is expected to be proportional to the thermal power difference, $\Delta S$, across the pn junction, as plotted in figure 4b. If the photocarrier transport is field driven, then the photocurrent is expected to follow the Fermi energy difference across the pn junction. We simulated the Fermi energy difference across the junction under different gate voltages using the split gate responses, and plotted it in figure 4c. The measured split gate voltage dependent photocurrent under 2 mW CW excitation is plotted in figure 4d. Clearly, the change of photocurrent polarity and peak show excellent agreement with simulation result of figure 4b. PTE dominates photocurrent generation in graphene pn junction, consistent with recent theoretical prediction based on similar device configuration.[18]

Last, we investigated the disappearance of photocurrent at the graphene pn junction under femtosecond pulse laser excitation, as previously shown in figure 1d. To exclude the effect of gate biasing, we reversed the gate biasing condition by sweeping $V_{g1}$ with $V_{g2}$ fixed at 0 V. Again, there is no photocurrent generation from pn junction (Figure 5a). A complete dual gate sweeps with pulse laser excitation at the pn junction show almost no photocurrent regardless of the dual gate voltages, as evident in figure 5b (also see Supporting Figure S3). This result further collaborates the hot carrier nature of the photocurrent generation in graphene. Under pulse excitation, there is no hot carrier temperature gradient across the pn junction due to the overheating phonon temperature[18] and PTE photocurrent is significantly suppressed as a result.

CONCLUSIONS

In summary, we systematically study the photocurrent generation at graphene-metal contact and graphene pn junction. The striking difference between CW and pulse laser excitation reveals that graphene photoresponse is closely related to illumination intensity. Importantly, we demonstrate the possibility of extracting non-equilibrium hot carriers from graphene. This finding may pave a promising pathway to build graphene based hot carrier optoelectronics. To improve the hot carrier extraction efficiency under low illumination intensity, hot carrier cooling rate could be further slowed down by quantizing graphene energy states through fabricating graphene nanoribbons,[26] or by opening bandgap in bilayer graphene.[27-28]

METHODS

Graphene was synthesised by chemical vapor deposition (CVD) method on copper[29], and then transferred to the pre-patterned substrate. Single layer nature of the graphene was identified by Raman spectroscopy. For the pre-patterned substrate, Ti/Au (5/30 nm) were patterned on top of 300 nm thick silicon dioxide to serve as two split bottom gates. The pair of split gates is separated by 1 μm. 50 nm thick $Al_2O_3$ was then deposited by atomic layer deposition (ALD) as the back gate dielectric. After transferring graphene to this pre-patterned substrate, the selected graphene channel areas were defined by oxygen plasma. Photolithography was then used to pattern source and drain contacts, and Ti/Au (5/50 nm) were deposited by electron beam evaporation. The source and drain contacts were separated by 5 μm. Finally, the entire graphene device was covered by 50nm $Al_2O_3$, deposited by ALD, in order to keep the device stable under ambient conditions.

**Conflict of Interest:** The authors declare no competing financial interest.

**Acknowledgements** We thank Profs. L. Jay Guo and P. C. Ku for sharing some of the equipments. This work was supported from the Donors of the American Chemical Society Petroleum Research Fund, the U-M/SJTU Collaborative Research Program in Renewable Energy Science and Technology, and National Science Foundation Scalable Nanomanufacturing Program (DMR-1120187). Devices were fabricated in the Lurie Nanofabrication Facility at University of Michigan, a member of the National Nanotechnology Infrastructure Network funded by the National Science Foundation.

**Supporting Information**

Detail device structure, electrical properties of the device, and scanning photocurrent spectroscopy. The material is available free of charge *via* the Internet at http://pubs.acs.org.

**Figure Captions**

**Figure 1 Striking difference for photocurrent generation in graphene between CW and pulse laser excitation. a,** Schematic drawing of the graphene device and experimental setup. **b,** Gate response of the device with $V_{sd} = 1$ mV. The black curve shows resistance dependence on $V_{g1}$ with $V_{g2}$ grounded, and the red curve shows $V_{g2}$ gate dependence with $V_{g1}$ grounded. The upper inset shows spatially resolved two dimensional photocurrent map with zero source-drain and gate bias voltage. The lower inset shows optical reflection intensity map of the same graphene device. The red dashed lines indicate the boundary of the source and drain contacts, and the green dashed lines indicate the boundary of split bottom gates. Scale bar, 2 µm. **c,** Gate dependent photocurrent map under 3.8 mW CW laser excitation. **d,** Gate dependent photocurrent map under 3.8 mW pulse laser excitation. In both 1c and 1d, the red dotted lines indicate the source and drain contacts edges, and the green dotted lines indicate the bottom gates edges.

**Figure 2 Extraction of photoexcited hot carriers from the graphene-metal junction. a,** Gate dependent photocurrent at the metal contact edge under CW laser excitation. **b,** Gate dependent photocurrent at the metal contact edge under pulse laser excitation. **c**, Schematic drawing of near-equilibrium carrier distribution under CW

laser excitation. **d**, Schematic drawing of nonequilibrium hot carrier distribution under pulse laser excitation.

**Figure 3 Power dependent hot carrier photocurrent. a, b, c,** Gate dependent photocurrent map under 580 μW (**a**), 930 μW (**b**), and 3.49 mW (**c**) pulse laser excitation, respectively. Position zero corresponds to the metal contact edge. **d,** Gate dependent photocurrent at the metal contact edge, excited by different pulse laser power. **e,** The zoom-in view of the low photocurrent amplitude region. The inset shows the relation between photocurrent peak and pulse laser power.

**Figure 4 Gate dependent photocurrent generation at graphene pn junction under CW excitation a,** Split gate responses of the device (lower panel) and the calculated gate dependent thermopower (upper panel). **b,** Thermopower difference across the pn junction under different split gate voltages. **c,** Fermi energy difference across the pn junction under different split gate voltages. **d,** Measured gate dependent photocurrent at graphene pn junction under 2 mW CW excitation.

**Figure 5 Gate dependent photocurrent generation at graphene pn junction under Pulse excitation a,** $V_{g1}$ gate dependent photocurrent map of the device under 2 mW pulse excitation. Here $V_{g2}$ is grounded. **b,** Dual gate dependent photocurrent map with pulse laser excitation at the graphene pn junction.

**REFERENCES AND NOTES**

1. Nair, R. R.; Blake, P.; Grigorenko, A. N.; Novoselov, K. S.; Booth, T. J.; Stauber, T.; Peres, N. M. R.; Geim, A. K. Fine Structure Constant Defines Visual Transparency of Graphene. *Science* **2008**, 320, 1308-1308.
2. Mak, K. F.; Sfeir, M. Y.; Wu, Y.; Lui, C. H.; Misewich, J. A.; Heinz, T. F. Measurement of the Optical Conductivity of Graphene. *Phys. Rev. Lett.* **2008**, 101, 196405.
3. Wang, H.; Strait, J. H.; George, P. A.; Shivaraman, S.; Shields, V. B.; Chandrashekhar, M.; Hwang, J.; Rana, F.; Spencer, M. G.; Ruiz-Vargas, C. S.; et al. Ultrafast Relaxation Dynamics of Hot Optical Phonons in Graphene. *Appl. Phys. Lett.* **2010**, 96, 081917.
4. Bistritzer, R.; MacDonald, A. H. Electronic Cooling in Graphene. *Phys. Rev. Lett.* **2009**, 102, 206410.
5. Strait, J. H.; Wang, H.; Shivaraman, S.; Shields, V.; Spencer, M.; Rana, F. Very Slow Cooling Dynamics of Photoexcited Carriers in Graphene Observed by Optical-Pump Terahertz-Probe Spectroscopy. *Nano Lett*. **2011**, 11, 4902-4906.
6. Ruzicka, B. A.; Wang, S.; Werake, L. K.; Weintrub, B.; Loh, K. P.; Zhao, H. Hot Carrier Diffusion in Graphene. *Phys. Rev. B* **2010**, 82, 195414.
7. Breusing, M.; Ropers, C.; Elsaesser, T. Ultrafast Carrier Dynamics in Graphite. *Phys. Rev. Lett.* **2009**, 102, 086809.
8. Breusing, M.; Kuehn, S.; Winzer, T.; Malic, E.; Milde, F.; Severin, N.; Rabe, J. P.; Ropers, C.; Knorr, A.; Elsaesser, T. Ultrafast Nonequilibrium Carrier Dynamics in a Single Graphene Layer. *Phys. Rev. B* **2011**, 83, 153410.
9. Dawlaty, J. M.; Shivaraman, S.; Chandrashekhar, M.; Rana, F.; Spencer, M. G. Measurement of Ultrafast Carrier Dynamics in Epitaxial Graphene. *Appl. Phys. Lett.* **2008,** 92, 042116.
10. Lui, C. H.; Mak, K. F.; Shan, J.; Heinz, T. F. Ultrafast Photoluminescence from Graphene. *Phys. Rev. Lett.* **2010**, 105, 127404.
11. Liu, W.-T.; Wu, S. W.; Schuck, P. J.; Salmeron, M.; Shen, Y. R.; Wang, F. Nonlinear Broadband Photoluminescence of Graphene Induced by Femtosecond Laser Irradiation. *Phys. Rev. B* **2010**, 82, 081408.
12. Lee, E. J. H.; Balasubramanian, K.; Weitz, R. T.; Burghard, M.; Kern, K. Contact and Edge Effects in Graphene Devices. *Nat. Nanotechnol.* **2008**, 3, 486-490.
13. Mueller, T.; Xia, F.; Freitag, M.; Tsang, J.; Avouris, P. Role of Contacts in Graphene Transistors: A Scanning Photocurrent Study. *Phys. Rev. B* **2009**, 79, 245430.

14. Xia, F.; Mueller, T.; Golizadeh-Mojarad, R.; Freitag, M.; Lin, Y.-M.; Tsang, J.; Perebeinos, V.; Avouris, P. Photocurrent Imaging and Efficient Photon Detection in a Graphene Transistor. *Nano Lett.* **2009**, 9, 1039-1044.
15. Park, J.; Ahn, Y. H.; Ruiz-Vargas, C. Imaging of Photocurrent Generation and Collection in Single-Layer Graphene. *Nano Lett.* **2009**, 9, 1742-1746.
16. Xu, X.; Gabor, N. M.; Alden, J. S.; van der Zande, A. M.; McEuen, P. L. Photo-Thermoelectric Effect at a Graphene Interface Junction. *Nano Lett.* **2010**, 10, 562-566.
17. Lemme, M. C.; Koppens, F. H. L.; Falk, A. L.; Rudner, M. S.; Park, H.; Levitov, L. S.; Marcus, C. M. Gate-Activated Photoresponse in a Graphene p-n Junction. *Nano Lett.* **2011**, 11, 4134-4137.
18. Song, J. C. W.; Rudner, M. S.; Marcus, C. M.; Levitov, L. S. Hot Carrier Transport and Photocurrent Response in Graphene. *Nano Lett.* **2011**, 11, 4688-4692.
19. Peters, E. C.; Lee, E. J. H.; Burghard, M.; Kern, K. Gate Dependent Photocurrents at a Graphene p-n Junction. *Appl. Phys. Lett.* **2010**, 97, 193102.
20. Gabor, N. M.; Song, J. C. W.; Ma, Q.; Nair, N. L.; Taychatanapat, T.; Watanabe, K.; Taniguchi, T.; Levitov, L. S.; Jarillo-Herrero, P. Hot Carrier-Assisted Intrinsic Photoresponse in Graphene. *Science* **2011**, 334, 648-652.
21. Sun, D.; Aivazian, G.; Jones, A. M.; Ross, J. S.; Yao, W.; Cobden, D.; Xu, X. Ultrafast Hot-Carrier-Dominated Photocurrent in Graphene. *Nat. Nanotechnol.* **2012**, 7, 114-118.
22. Calandra, M.; Mauri, F. Electron-Phonon Coupling and Electron Self-Energy in Electron-Doped Graphene: Calculation of Angular-Resolved Photoemission Spectra. *Phys. Rev. B* **2007**, 76, 205411.
23. Tse, W.-K.; Das Sarma, S. Phonon-Induced Many-Body Renormalization of the Electronic Properties of Graphene. *Phys. Rev. Lett.* **2007**, 99, 236802.
24. Tse, W.-K.; Hwang, E. H.; Sarma, S. D. Ballistic Hot Electron Transport in Graphene. *Appl. Phys. Lett.* **2008**, 93, 023128.
25. Winzer, T.; Knorr, A.; Malic, E. Carrier Multiplication in Graphene. *Nano Lett.* **2010**, 10, 4839-4843.
26. Li, X.; Wang, X.; Zhang, L.; Lee, S.; Dai, H. Chemically Derived, Ultrasmooth Graphene Nanoribbon Semiconductors. Science 2008, 319, 1229-1232.
27. Zhang, Y.; Tang, T.-T.; Girit, C.; Hao, Z.; Martin, M. C.; Zettl, A.; Crommie, M. F.; Shen, Y. R.; Wang, F. Direct Observation of a Widely Tunable Bandgap in Bilayer Graphene. *Nature* **2009**, 459, 820-823.
28. Lee, S.; Lee, K.; Zhong, Z. Wafer Scale Homogeneous Bilayer Graphene Films by Chemical Vapor Deposition. *Nano Lett.* **2010**, 10, 4702-4707.

29. Li, X.; Cai, W.; An, J.; Kim, S.; Nah, J.; Yang, D.; Piner, R.; Velamakanni, A.; Jung, I.; Tutuc, E.; et al. Large-Area Synthesis of High-Quality and Uniform Graphene Films on Copper Foils. *Science* **2009**, 324, 1312-134.

**Figures**

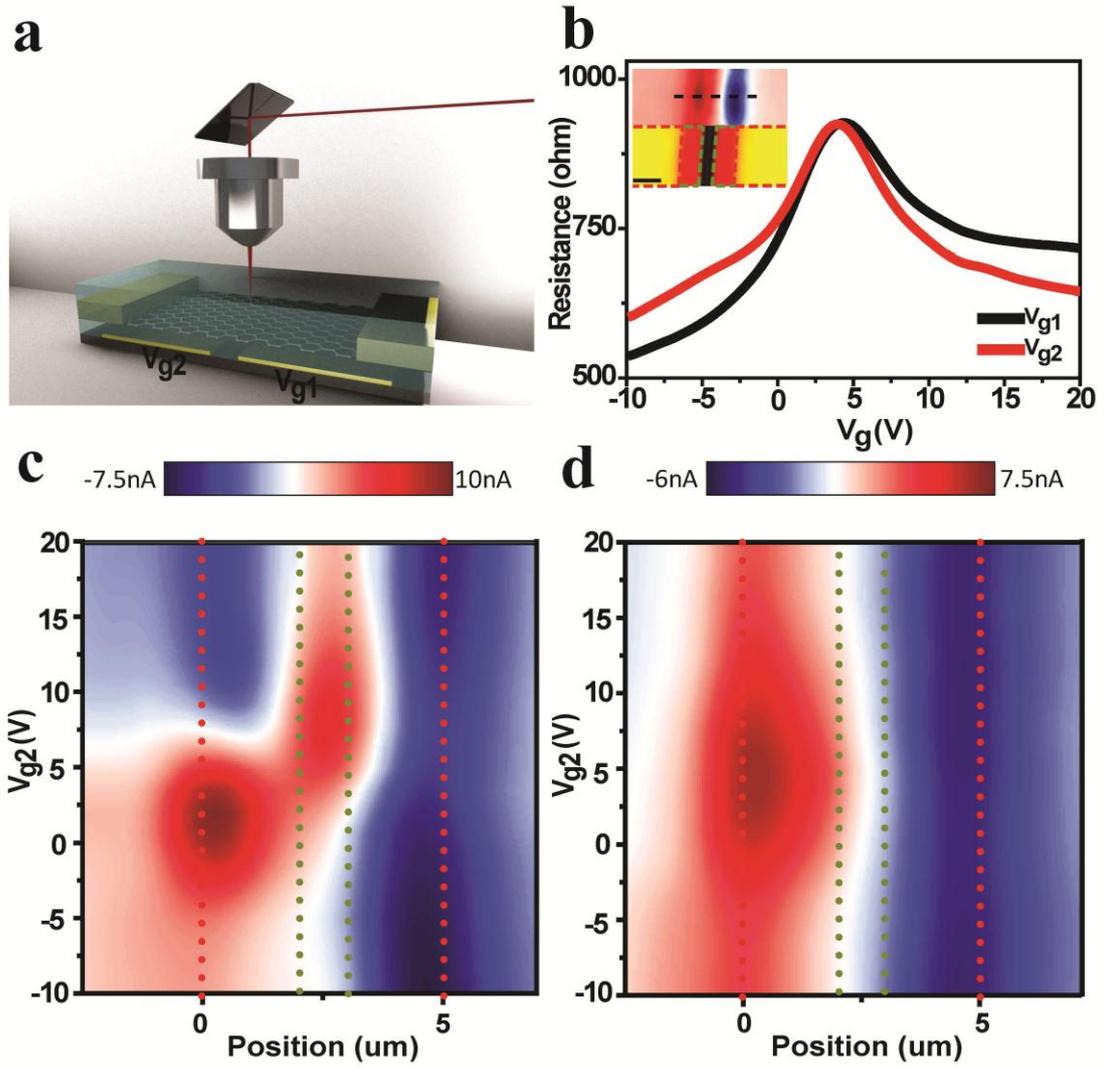

**Figure 1**

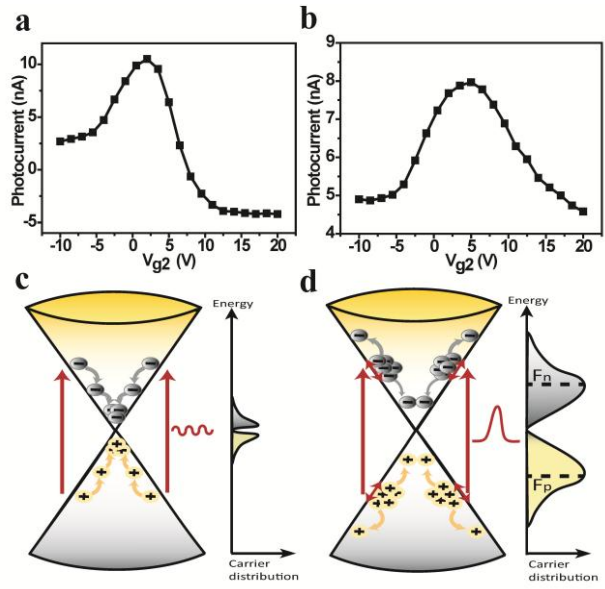

**Figure 2**

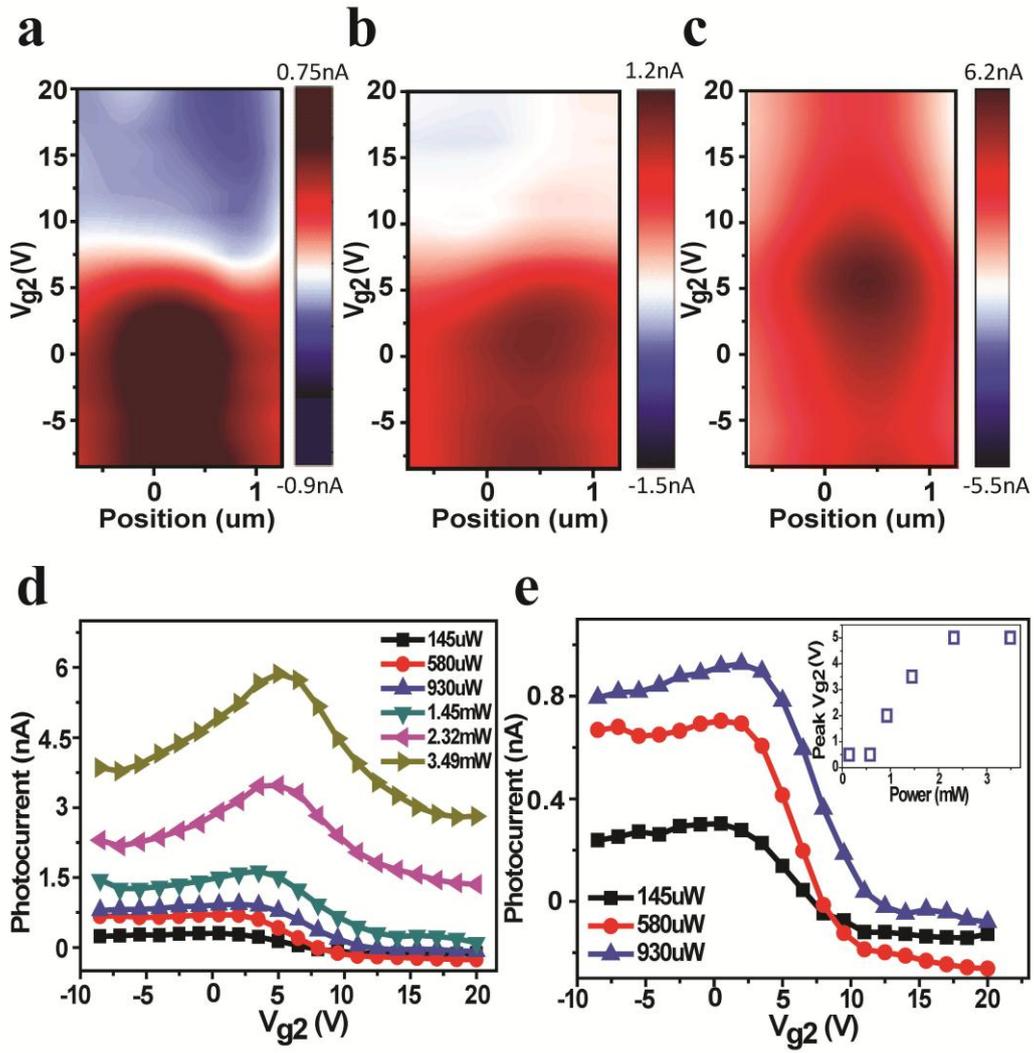

**Figure 3**

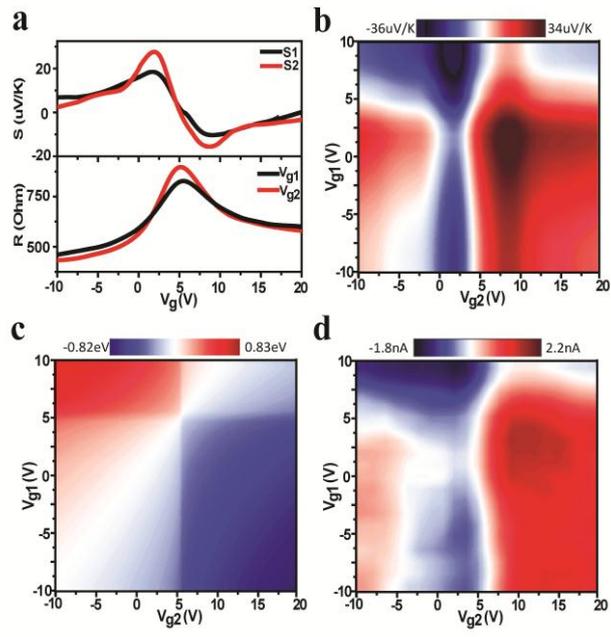

**Figure 4**

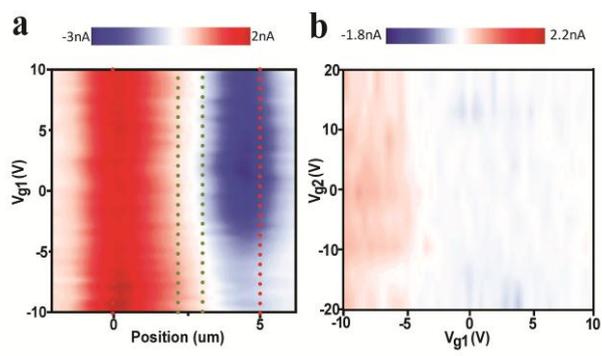

**Figure 5**

**TOC Graphic**

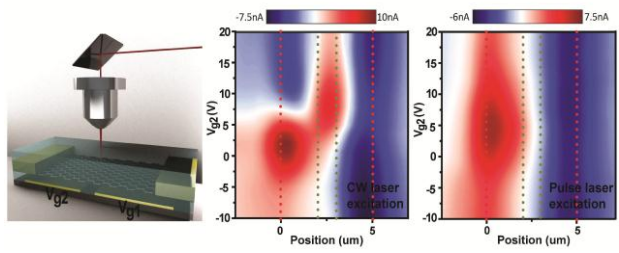